\documentclass[aps, prd,   showpacs, twocolumn]{revtex4}

\usepackage{amsmath,amsfonts,amscd,amssymb,epsf}
\usepackage[small]{caption}

\newcommand{\pa}{\partial}

\begin{document}

\title{ Casimir force in noncommutative Randall-Sundrum models revisited}

 \author{L.P. Teo}\email{ LeePeng.Teo@nottingham.edu.my}\address{Department of Applied Mathematics, Faculty of Engineering, University of Nottingham Malaysia Campus, Jalan Broga, 43500, Semenyih, Selangor Darul Ehsan, Malysia. }

\begin{abstract}
We propose another method to compute the Casimir force in noncommutative Randall-Sundrum braneworld model considered by K. Nouicer and Y. Sabri  recently.  Our method can be used to compute  the Casimir force to any order    in the noncommutative parameter.  Contrary to the claim made by K. Nouicer and Y. Sabri  that repulsive Casimir force can appear in the first order approximation,   we show that the Casimir force is always attractive at any order of approximation.
\end{abstract}

\pacs{ 11.25.Mj, 11.10.Kk, 11.10.Nx}

 \maketitle

In a recent publication \cite{1}, K. Nouicer and Y. Sabri computed the Casimir force acting on a pair of parallel plates in the five dimensional Randall-Sundrum braneworlds of type I and type II. They computed the Casimir force to the first order in the noncommutative parameter and claimed that in Randall-Sundrum model of type I (RSI), the presence of noncommutativity will lead to repulsive Casimir force when   the plate separation is small.

In this report, we   present another   computation method that allows us to compute to any order of the noncommutative parameter, and investigate whether noncommutativity
will change the nature of the Casimir force. Let $\kappa$ be the parameter governing the degree of curvature of the RSI model. Define $\xi=\pi \kappa e^{-\pi \kappa R}$ and let $l$ be the fundamental noncommutative length scale. The Casimir energy between a pair of parallel plates with distance $a$ apart   is given by
\begin{equation}\label{eq4_1_1}\begin{split}
E_{\text{Cas}}(a)= \frac{A\hbar c }{2} \int_{\mathbb{R}^2}&\frac{d^2\boldsymbol{k}_{\perp}}{(2\pi)^2} \left(  p'\sum_{n=1}^{\infty}\omega_{n0}e^{- l^2\omega_{n0}^2 }
\right.\\&\left. +v  p\sum_{N=1}^{\infty}\sum_{n=1}^{\infty} \omega_{nN}e^{- l^2\omega_{nN}^2 } \right),\end{split}
\end{equation}where $A$ is the area of the plates and $v=2$ accounts for the volume of the orbifold \cite{4}. The eigenfrequencies $\omega_{nN}$ are given by
\begin{equation*}
\begin{split}
\omega_{nN}=&   \sqrt{|\boldsymbol{k}_{\perp}|^2+\left(\frac{\pi n}{a}\right)^2+ \kappa_N^2},
\end{split}\end{equation*}where $\kappa_N, N=0,1,2,\ldots,$ are the effective masses due to the existence of the extra dimension, and $\kappa_0=0$. We put the factors $p'$ and $p$ in \eqref{eq4_1_1} so that we can compare to  \cite{1}. If one considers massless scalar field with Dirichlet boundary conditions, one should take $p=p'=1$. In \cite{1}, it was claimed that for electromagnetic field with perfectly conducting boundary conditions, one should set $p'=2$ and $p=3$  due to the polarizations of photons in 4D and 5D spacetime. For Casimir effect in extra dimensional spacetime as in the present scenario, it is actually questionable whether one can obtain the result for electromagnetic field by simply adding the polarization factors $p'$ and $p$.  Such an imposition of polarization factors for Casimir effect in spacetime with extra dimensions was first applied in the paper \cite{5}, and was later followed by other works such as \cite{2,6,7}.  However, the Casimir effect of electromagnetic field in spacetime with extra dimensions is actually not that simple, and it highly depends on the geometry of the extra dimensions and on how one interprets the perfectly conducting boundary conditions. Different approaches will lead to different results, and one may not be able to obtain the correct 4D limit when the size of the extra dimensions vanishes.  Therefore majority of the works in Casimir effect in spacetime with extra dimensions work with scalar field rather than electromagnetic field.  Nevertheless, it should be pointed out that in the recent work \cite{9}, the authors found a new approach to the Casimir effect of electromagnetic field in 5D Kaluza Klein spacetime. They treated the Kaluza-Klein excitations of the electromagnetic field as Proca fields for massive photons, and apply the result of \cite{10} on Casimir effect of massive photons. In this approach, one find two discrete and one continuous longitudinal polarizations for the photons. In the limit the extra dimension diminishes, the contributions from the two discrete polarizations yield the conventional 4D result, and the contribution from the continuous longitudinal polarization vanishes. One can use the same approach for electromagnetic field in Randall-Sundrum spacetime. However, this will take us too far afield. Since our concern here is whether noncommutativity will change the sign of the Casimir force, it is sufficient to consider massless scalar field with Dirichlet boundary condition.

Now we go back to the computation of the Casimir energy. Notice that for any $l>0$, the Casimir energy \eqref{eq4_1_1} is finite. As $l\rightarrow 0$, it gives the Casimir energy in the absence of noncommutativity which is divergent in its naive definition. Since we would like to find the small $l$-expansion of the Casimir energy,  we regularize the Casimir energy as in the zeta regularization:
\begin{equation}\label{eq4_1_3}\begin{split}
E_{\text{Cas}}(a)= \frac{A\hbar c \mu^{2s}}{2}& \int_{\mathbb{R}^2}\frac{d^2\boldsymbol{k}_{\perp}}{(2\pi)^2} \left( p'\sum_{n=1}^{\infty}\omega_{n0}^{1-2s}e^{-l^2\omega_{n0}^2}
\right.\\&\left.\left. +v p\sum_{N=1}^{\infty}\sum_{n=1}^{\infty} \omega_{nN}^{1-2s}e^{-l^2\omega_{nN}^2} \right)\right|_{s=0}.\end{split}
\end{equation}Here $\mu$ is a normalization parameter.

Let us first consider the term
\begin{equation}\label{eq4_1_2}
 \left. \frac{A\hbar cp'  \mu^{2s}}{2} \int_{\mathbb{R}^2}\frac{d^2\boldsymbol{k}_{\perp}}{(2\pi)^2}   \sum_{n=1}^{\infty}\omega_{n0}^{1-2s}e^{-l^2\omega_{n0}^2}\right|_{s=0}
\end{equation}which is the Casimir energy in the absence of the extra dimension. Expanding  the exponential term  gives
\begin{equation*}
\begin{split}
&  \frac{A\hbar  cp' \mu^{2s}}{2} \int_{\mathbb{R}^2}\frac{d^2\boldsymbol{k}_{\perp}}{(2\pi)^2}    \sum_{n=1}^{\infty}\omega_{n0}^{1-2s}e^{-l^2\omega_{n0}^2}\\
=&\frac{A\hbar c p' \mu^{2s}}{4\pi}\int_0^{\infty} k dk \sum_{j=0}^{\infty}\frac{(-1)^jl^{2j} }{j!}\sum_{n=1}^{\infty} \left(k^2+\left(\frac{\pi n}{a}\right)^2\right)^{j+\frac{1}{2}-s}\\
=&-\frac{A\hbar  c p'\mu^{2s}}{8\pi} \sum_{j=0}^{\infty}\frac{(-1)^jl^{2j} }{j!}\left(\frac{\pi}{a}\right)^{2j+3-2s}\frac{\zeta_R(2s-2j-3)}{j+\frac{3}{2}-s},
\end{split}
\end{equation*}where $\zeta_R(z)$ is the Riemann zeta function. Setting $s=0$, we find that the contribution to the Casimir energy   in the absence of the extra dimension is
\begin{equation}\label{eq4_1_6}
\begin{split}
-\frac{A\hbar cp'}{8\pi} \sum_{j=0}^{\infty}\frac{(-1)^jl^{2j} }{j!}\left(\frac{\pi}{a}\right)^{2j+3}\frac{\zeta_R(-2j-3)}{j+\frac{3}{2}}.
\end{split}
\end{equation}
For the second term in \eqref{eq4_1_3}, similar computation gives
\begin{equation}\label{eq4_1_5}
\begin{split}
&\frac{A\hbar cvp \mu^{2s}}{2}\int_{\mathbb{R}^2}\frac{d^2\boldsymbol{k}_{\perp}}{(2\pi)^2}   \sum_{N=1}^{\infty}\sum_{n=1}^{\infty} \omega_{nN}^{1-2s}e^{-l^2\omega_{nN}^2} \\
=&-\frac{A\hbar cvp \mu^{2s}}{8\pi} \sum_{j=0}^{\infty}\frac{(-1)^jl^{2j} }{j!} \frac{Z\left(s-j-\tfrac{3}{2}\right)}{j+\frac{3}{2}-s},
\end{split}
\end{equation}where
\begin{equation*}
\begin{split}
&Z(s)=\sum_{N=1}^{\infty}\sum_{n=1}^{\infty} \left( \left(\frac{\pi n}{a}\right)^2+\kappa_N^2\right)^{ -s}\\
=&\frac{1}{\Gamma\left(s \right)}\int_0^{\infty} t^{s-1 }\sum_{N=1}^{\infty}\sum_{n=1}^{\infty}\exp\left\{-t \left( \left(\frac{\pi n}{a}\right)^2+\kappa_N^2\right)\right\}dt.
\end{split}
\end{equation*}Using the    formula
\begin{equation*}
\sum_{n=-\infty}^{\infty} \exp\left(-\alpha n^2\right)=\sqrt{\frac{\pi}{\alpha}}\sum_{n=-\infty}^{\infty} \exp\left( -\frac{\pi^2 n^2}{\alpha}\right),
\end{equation*}we find that
\begin{equation}\label{eq4_1_4}
\begin{split}
&Z(s)=-\frac{1}{2}\frac{1}{\Gamma\left(s \right)}\int_0^{\infty} t^{s-1}\sum_{N=1}^{\infty} e^{-t  \kappa_N^2 }dt\\
&+\frac{a}{2}\frac{1}{\sqrt{\pi}\Gamma(s)}\int_0^{\infty} t^{s-\frac{3}{2}}\sum_{N=1}^{\infty} e^{-t  \kappa_N^2 }dt\\
&+\frac{a}{\sqrt{\pi}}\frac{1}{\Gamma(s)}\int_0^{\infty} t^{s-\frac{3}{2}}\sum_{N=1}^{\infty} \sum_{n=1}^{\infty} e^{-\frac{a^2 n^2}{t }-t\kappa_N^2}dt\\
=&-\frac{1}{2}\zeta_{M}(s) +\frac{a}{2\sqrt{\pi}}\frac{\Gamma\left(s-\frac{1}{2}\right)}{\Gamma(s)}\zeta_M\left(s-\frac{1}{2}\right)\\
&+\frac{2a}{\sqrt{\pi}\Gamma(s)}\sum_{n=1}^{\infty}\sum_{N=1}^{\infty} \left(\frac{an}{\kappa_N}\right)^{s-\frac{1}{2}}K_{s-\frac{1}{2}}\left(2an\kappa_N\right),
\end{split}
\end{equation}where $K_{\nu}(z)$ is the modified Bessel function of second kind and
$$\zeta_M(s)=\sum_{N=1}^{\infty}\kappa_N^{-2s}.$$ Substituting the result of \eqref{eq4_1_4} into \eqref{eq4_1_5}, we find that the second term of \eqref{eq4_1_3} is equal to
\begin{equation}\label{eq4_1_7}
\begin{split}
&-\frac{A\hbar c v p}{8\pi } \sum_{j=0}^{\infty}\frac{(-1)^jl^{2j} }{j!}\left\{-\left.\frac{\mu^{2s}}{2}\frac{\zeta_M\left(s-j-\tfrac{3}{2}\right)}{ j+\frac{3}{2}-s}\right|_{s=0}\right.\\
& +\left.\frac{a}{2\sqrt{\pi}}\mu^{2s}\frac{\Gamma(s-j-2)}{\Gamma\left(s-j-\frac{3}{2}\right)}\frac{\zeta_M\left(s-j-2\right)}{j+\frac{3}{2}-s}\right|_{s=0}\\
&+  \frac{2a}{\left(j+\frac{3}{2}\right)\sqrt{\pi}\Gamma\left(-j-\frac{3}{2}\right)}\\&\times \left.\sum_{n=1}^{\infty}\sum_{N=1}^{\infty} \left(\frac{\kappa_N}{an}\right)^{j+2}K_{j+2}\left(2an\kappa_N\right)\right\}.
\end{split}
\end{equation}The first two terms might contain divergences when we set $s=0$. The sum of the terms \eqref{eq4_1_6} and \eqref{eq4_1_7} gives the Casimir energy between the plates.  Now to investigate the effect of Casimir energy on the plates, one need to take into account the contribution from the outside of the plates. This can be achieved by the so-called piston approach \cite{3}, where one considers a system of three plates at $x=0, x=a$ and $x=L$ respectively. In the limit $L\rightarrow \infty$,  the chamber between $x=a$ and $x=L$ is considered as the 'outside' of the plate at $x=a$. The   Casimir energy of the parallel plate system is defined as
$$E_{\text{Cas}}^{\parallel}(a)=\lim_{L\rightarrow \infty}\left(E_{\text{Cas}}(a)+E_{\text{Cas}}(L-a)-E_{\text{Cas}}(L)\right),$$   which is the sum of the Casimir energies of the two chambers divided by the plate at $x=a$, minus the Casimir energy in the absence of plates.  From \eqref{eq4_1_6} and \eqref{eq4_1_7}, we find that the Casimir energy of the parallel plate system is given by
\begin{equation}\label{eq4_1_8}
\begin{split}
&E_{\text{Cas}}^{\parallel}(a)=-\frac{A\hbar c p'}{8\pi} \sum_{j=0}^{\infty}\frac{(-1)^jl^{2j} }{j!}\left(\frac{\pi}{a}\right)^{2j+3}\frac{\zeta_R(-2j-3)}{j+\frac{3}{2}}\\
&-\frac{A\hbar c v p}{8\pi }  \sum_{j=0}^{\infty}\frac{(-1)^jl^{2j} }{j!}\left\{-\left.\frac{\mu^{2s}}{2}\frac{\zeta_M\left(s-j-\tfrac{3}{2}\right)}{ j+\frac{3}{2}-s}\right|_{s=0}\right.\\
&+  \frac{2a}{\left(j+\frac{3}{2}\right)\sqrt{\pi}\Gamma\left(-j-\frac{3}{2}\right)}\\&\times \left.\sum_{n=1}^{\infty}\sum_{N=1}^{\infty} \left(\frac{\kappa_N}{an}\right)^{j+2}K_{j+2}\left(2an\kappa_N\right)\right\}.
\end{split}
\end{equation}
The Casimir force per unit area acting on the parallel plates is given by
\begin{equation*}
\begin{split}
\mathcal{F}_{\text{Cas}}^{\parallel}(a)=-\frac{1}{A}\frac{\pa E_{\text{Cas}}^{\parallel}(a)}{\pa a}.
\end{split}
\end{equation*}Using the formula
$$K_{\nu}'(z)=-\frac{1}{2}\left(K_{\nu-1}(z)+K_{\nu+1}(z)\right),$$ we find from \eqref{eq4_1_8} that to all orders in the noncommutative length parameter $l$,
\begin{equation}\label{eq4_1_10}
\begin{split}
&\mathcal{F}_{\text{Cas}}^{\parallel}(a)=-\frac{\hbar cp'}{4\pi }\sum_{j=0}^{\infty}\frac{(-1)^jl^{2j}}{j!}\frac{\pi^{2j+3}}{a^{2j+4}}\zeta_R(-2j-3)\\
&-\frac{\hbar cvp}{4\pi^{\frac{3}{2}}}\sum_{j=0}^{\infty}\frac{(-1)^j l^{2j}}{j!\left(j+\frac{3}{2}\right) \Gamma\left(-j-\frac{3}{2}\right)} \\&\times\left\{\frac{1}{a^{j+2}}\sum_{n=1}^{\infty}\sum_{N=1}^{\infty} (j+1) \left(\frac{\kappa_N}{ n}\right)^{j+2}K_{j+2}(2an\kappa_N)\right.\\
&+\left.\frac{1}{a^{j+1}}\sum_{n=1}^{\infty}\sum_{N=1}^{\infty} \frac{\kappa_N^{j+3}}{n^{j+1}}\left(K_{j+1}(2an\kappa_N)+K_{j+3}(2an\kappa_N)\right)\right\}.
\end{split}
\end{equation}
Using the functional equation $$\pi^{-\frac{s}{2}}\Gamma\left(\frac{s}{2}\right)\zeta_R(s)=\pi^{-\frac{1-s}{2}}\Gamma\left(\frac{1-s}{2}\right)\zeta_R(1-s),$$and the formulas
 $$s\Gamma(s)=\Gamma(s+1), \hspace{1cm}
 \Gamma(s)\Gamma(1-s)=\frac{\pi}{\sin(\pi s)},$$ we find that \eqref{eq4_1_10} can be rewritten as
\begin{widetext}\begin{equation}\label{eq4_1_11}
\begin{split}
&\mathcal{F}_{\text{Cas}}^{\parallel}(a)=-\frac{\hbar cp'}{4\pi^{\frac{5}{2}}}\sum_{j=0}^{\infty}l^{2j}(j+1)\Gamma\left(j+\frac{5}{2}\right)\zeta_R(2j+4)-\frac{\hbar cvp}{4\pi^{\frac{5}{2}}}\sum_{j=0}^{\infty}l^{2j}\frac{\Gamma\left(j+\frac{3}{2}\right)}{j!}\\&\times\left\{\frac{1}{a^{j+2}}\sum_{n=1}^{\infty}\sum_{N=1}^{\infty} (j+1) \left(\frac{\kappa_N}{ n}\right)^{j+2}K_{j+2}(2an\kappa_N)\right.+\left.\frac{1}{a^{j+1}}\sum_{n=1}^{\infty}\sum_{N=1}^{\infty} \frac{\kappa_N^{j+3}}{n^{j+1}}\left(K_{j+1}(2an\kappa_N)+K_{j+3}(2an\kappa_N)\right)\right\}.
\end{split}
\end{equation}\end{widetext}Since for $z>0$, $K_{\nu}(z)$ and $\Gamma(z)$ are positive, and $\zeta_R(z)$ is positive for all $z>1$, we see from \eqref{eq4_1_11} that in each order of  the noncommutative parameter $l$, the Casimir force is always negative (attractive).

To compare with the result of \cite{1}, let us compute the first order approximation of the Casimir force density.   From \eqref{eq4_1_10}, we find that with the approximation $\displaystyle \kappa_N\approx \xi\left(N+\tfrac{1}{4}\right)$ used in \cite{1}, the sum of the $j=0$ and $j=1$ terms is
\begin{widetext}
\begin{equation}\label{eq4_2_1}\begin{split}
&\mathcal{F}_{\text{Cas}}^{\parallel}(a)\approx - \frac{\hbar c p' \pi^2}{4a^4}\left(\zeta_R(-3)- l^2  \left(\frac{\pi}{a}\right)^{2}\zeta_R(-5)\right)
-\frac{\hbar c p \xi^2}{4\pi^2 a^2}\left\{   \sum_{n=1}^{\infty}\sum_{N=1}^{\infty}\frac{\left(N+\tfrac{1}{4}\right)^2}{n^2}K_2\left(2\xi a n \left(N+\frac{1}{4}\right)\right)\right.\\&\left. +\xi a\sum_{n=1}^{\infty} \sum_{N=1}^{\infty}\frac{\left(N+\tfrac{1}{4}\right)^3}{n}
\left(K_1\left(2\xi a n\left(N+\tfrac{1}{4}\right)\right)+K_3\left(2\xi a n\left(N+\tfrac{1}{4}\right)\right)\right)\right\}-\frac{3l^2\hbar c p \xi^3}{8\pi^2 a^3}\left\{  2 \sum_{n=1}^{\infty}\sum_{N=1}^{\infty}\frac{\left(N+\frac{1}{4}\right)^3}{n^3}K_3\left(2\xi a n \left(N+\tfrac{1}{4}\right)\right)\right.\\&\left. +\xi a\sum_{n=1}^{\infty} \sum_{N=1}^{\infty}\frac{\left(N+\tfrac{1}{4}\right)^4}{n^2}
\left(K_2\left(2\xi a n\left(N+\tfrac{1}{4}\right)\right)+K_4\left(2\xi a n\left(N+\tfrac{1}{4}\right)\right)\right)\right\}.
\end{split}\end{equation}
\end{widetext}The first term \begin{equation}\label{eq4_2_3}- \frac{\hbar c p' \pi^2}{4a^4}\left(\zeta_R(-3)- l^2  \left(\frac{\pi}{a}\right)^{2}\zeta_R(-5)\right) \end{equation}gives the Casimir force density in the absence of the extra dimension. Using $\zeta(-3)=1/120, \zeta(-5)=-1/252$, we find that it is equal to \begin{equation}\label{eq4_2_3} -\frac{\hbar c p'  \pi^2}{480 a^4}-\frac{\hbar c p'  \pi^4 l^2}{1008 a^6}.\end{equation}
Compare \eqref{eq4_2_1} to eq. (45) in \cite{1}, we find that everything agrees except for the sign of the term $\displaystyle -\frac{3l^2\hbar c p \xi^3}{8\pi^2 a^3}$  in front of the last pair of big brackets in \eqref{eq4_2_1}, which we find to be negative but was put as positive in \cite{1}. We follow the derivation of eq. (45) in \cite{1} and find that this discrepancy is due to a sign error made by the authors of \cite{1}. This sign error in \cite{1} has lead to the wrong conclusion that the Casimir force can become repulsive, in contrary to our result that the Casimir force is always attractive.

In conclusion, we have proposed another method for computing the Casimir force in noncommutative spacetime. We find that there is an error in the computations presented in \cite{1} which leads to incorrect conclusion that the Casimir force can become repulsive.  Although we have worked in the specific model that are considered in \cite{1}, our method actually can be used to for other models.
As mentioned in the beginning of this report, replacing $p'=p=1$ by $p'=2$ and $p=3$ might not be the correct approach for passing from massless scalar field to electromagnetic field. Using the approach of \cite{9} for Casimir effect of electromagnetic field in spacetime with extra dimensions, one actually has two discrete modes and a continuous longitudinal mode. The contribution from the discrete modes   is the same as taking $p'=p=2$ in this report, but the contribution from the longitudinal modes can be considered as a special case of dielectrics, for which one can apply Lifshitz formula. In the absence of noncommutativity, it has been verified in \cite{9} that the contribution from the longitudinal modes is much weaker than the contribution from the discrete modes, but both of them are attractive. It is not obvious that noncommutativity will not change the sign of the longitudinal contribution. But the method discussed in this report can be used to prove that this is indeed the case. We shall discuss this in more detail elsewhere.
We conclude that in general,  noncommutativity of spacetime will not alter the attractive nature of the Casimir force acting between parallel plates.

\begin{acknowledgments}
We would like to thank the authors of \cite{1} and the anonymous referee whose comments help to improve this manuscript.
\end{acknowledgments}

\end{document}